\documentstyle[pre,aps]{revtex}
\begin{document}
\draft
\title{Optimized random phase approximations for arbitrary reference 
systems: \\ extremum conditions and thermodynamic consistence}

\author{G. Pastore\\{\it Istituto Nazionale di Fisica della Materia 
}\\{\it
and  Dipartimento di Fisica Teorica dell' Universit\`{a}}\\
{\it Strada Costiera 11, 34014 Trieste, Italy}\\
O. Akinlade\\ {\it Department of Physics, University of Agriculture},
\\ {\it Abeokuta, Nigeria}\\
{\it and International Centre for Theoretical Physics, Trieste, Italy}\\
F. Matthews\\{\it Department of Physics, Ondo State University},
\\ {\it Ado Ekiti,
Nigeria}\\{\it and International Centre for Theoretical Physics, Trieste, Italy}
\\Z. Badirkhan\\{\it International Centre for Theoretical Physics, Trieste, Italy}}
\date{\today}
\maketitle

\begin{abstract}
The optimized random phase approximation (ORPA) for classical liquids
is re-examined in the framework of the generating functional approach 
to the integral equations.
We show that the two main variants of the approximation 
correspond to the addition of the same  correction to two different 
first order approximations of the homogeneous liquid free energy. 
Furthermore, we  show that it is possible to consistently use the ORPA
with arbitrary reference systems described by continuous potentials 
and that  the same 
approximation is equivalent to a particular extremum condition for the 
corresponding
generating functional. Finally, it is possible to enforce the 
thermodynamic consistence
between the thermal and the virial route to the equation of state by 
requiring the global extremum condition on the generating functional.
\vskip .2truecm

\end{abstract}

\section{Introduction}

The optimized random phase approximation (ORPA)\cite{WCA} 
for classical liquids has been
extensively used in the last decades to obtain information on the structure and,
to a minor extent, on the thermodynamics of simple liquids and mostly  
liquid metals \cite{WCA,Regnault,ACW,LJ,KH,PT}.
The approximation was originally developed in the context of the  perturbative approach
to the thermodynamic and structure of simple liquids\cite{WCA}. 
More recently, this approximation has been used as ingredient for studying
liquids in porous media\cite{porous} and critical phenomena in simple
liquids\cite{Reatto}. The standard
implementation of ORPA is based on the splitting of the interparticle potential
into a repulsive (reference)  and an attractive (perturbation) part. 
The effect of the attraction on the
pair correlation functions of the purely repulsive reference system is treated
at the level of the random phase approximation (RPA) at large distances and by
enforcing the excluded volume effect at short distances.

Numerical studies have shown that the ORPA yields a very accurate description of the
structure factor and  thermodynamics of simple fluids. At  present, for simple
liquids, the
quality of the ORPA results is comparable to that of 
state-of-the-art calculations based on 
the modified hypernetted
chain (MHNC) approximation\cite{RA,ladovar} or other modern integral equations 
like the HMSA\cite{ZH}.

Usually, the reference system is modeled by hard sphere interactions. 
However, in some cases, either the nature of the interactions or
results from other theoretical  approaches, indicate that a soft reference system
could provide a better  reference system  for the ORPA. 
The original derivation of the ORPA does not allow a direct extension
of the formulae to the case of a reference system  interacting 
without hard core. A non-standard
implementation for liquid alkali metals using a one component plasma as 
reference system\cite{PT}, although providing good results, was not 
general enough and was subject to some criticisms\cite{Hafner}.

More recently, motivated by the need of improving some variational calculations for 
liquid
metals\cite{GBmetal}, we reviewed the ORPA from the computational as well
from the theoretical point of view. In a previous paper\cite{ORPA1},
we showed that the solution of the ORPA equations is unique and 
we proposed a new robust,
accurate and efficient numerical algorithm to solve the equations. In the present
paper we address the problem of reformulating  the theory in order to 
deal in a consistent way with continuous reference systems 
without any intermediate 
introduction of auxiliary hard-sphere systems\cite{WCA}.
We give a solution to this problem  in the same spirit of Gillan's 
extension of the 
mean spherical approximation for soft potentials\cite{gillan} and subsequent
elaborations by Rosenfeld\cite{ros_msa}. 
 
We found useful to  recast the ORPA equations in the framework of the generating
functionals for the integral equations of the theory of liquids\cite{Olivares}.
We show that the two different first-order expansions of the free 
energy  functional corresponding to the well known Gibbs-Bogolioubov and 
Weeks-Chandler and Andersen approximations for the free energy\cite{HMD}
can be transformed into two closely related forms of the ORPA by addition of the 
same functional. The resulting correlation functions differ by the  choice
of the reference system  pair correlation function.

Moreover, we are  able to show that our 
condition for a continuous ORPA correction to the pair correlations is 
equivalent
to an extremum condition for the ORPA generating functional. Since only
the variational determination of all the free parameters of the 
functional allows the identification of its value  
with the Helmoltz free energy, this choice enforces the thermodynamic 
consistence in a natural way.

The paper is organized as follows. In section 2, we show how
two versions of the ORPA differing only in  the treatment of the
reference system correlations can be obtained from a variational 
problem for two related  functionals.  In section 3 we show that it 
is possible to define a consistent ORPA for continuous potentials and that such an
extension is equivalent to an additional requirement of  extremum for the ORPA
functionals.
In section 4 the issue of the thermodynamic
consistence of the two approximations is briefly discussed.
Conclusions are summarized in section 5.

\section{Two generating functionals for the  ORPA}

The starting point of the ORPA is a suitable decomposition of the
interatomic potential $\phi(r)$ into a reference potential $\phi_0(r)$ and a
perturbation (the rest) $\phi_1(r)$:

\begin{equation}
\phi(r) = \phi_0(r) + \phi_1(r) \label{pot}
\end{equation}

Although the original ORPA \cite{WCA} was based on a specific choice of such a
decomposition, in the following discussion we temporarily leave unspecified the 
exact
characterization of $\phi_0(r)$. Equation (1)
naturally leads to a similar
decomposition of the total and the direct correlation functions $h(r)$ 
and $c(r)$:

\begin{eqnarray}
h(r) & = & h_0(r) + \Delta h(r) \label{hdec}\\
c(r) & = & c_0(r) + \Delta c(r) \label{cdec}
\end{eqnarray}

where $h_0(r)$ and $c_0(r)$ are the correlation functions of a reference fluid
whose particles interact via the potential $\phi_0(r)$. The 
thermodynamics and the  
correlation functions of the reference system are considered as known 
quantities. $\Delta h(r)$
and $\Delta c(r)$ are defined  by equations (2) and (3)
and are the unknown
functions of the theory. A relation among them, for a fluid whose 
number density is $\rho$, is provided by the 
Ornstein-Zernike equation

\begin{equation}
h(r) = c(r) + \rho \int {\rm d}^3{\bf r'} h(r') c( \left| {\bf r - r' } \right| )
\label{OZ}
\end{equation}

which, taking into account the fact that $h_0(r)$ and $c_0(r)$ do satisfy the 
same
equation, results in the following relation between the Fourier transforms of 
$\Delta h(r)$ and $ \Delta c(r)$:

\begin{equation}
\Delta \hat{h} (q) = 
\frac{  \Delta \hat{c} (q) S^2_0(q)}{  1 - \rho \Delta \hat{c}(q) S_0(q) }
\label{SRPA}
\end{equation}

In formula (\ref{SRPA}), $S_0(q)= 1 + \rho \hat{h_{0}}(q)$ is the structure factor of
the reference system. A hat on a function of $q$ 
indicates the three-dimensional 
Fourier transforms of the corresponding function defined in the $r$-space 
while 
$\rho$ is the number density of the system.

So far no approximation has been introduced yet. By complementing equation 
(\ref{SRPA})
with any approximate relation between $\Delta h(r),\Delta c(r)$ and 
$ \phi_1(r)$ 
we get a closed set of non-linear integral equations that has to be solved.

In particular, the ORPA closure corresponds to  the dual relations:

\begin{eqnarray}
\Delta c(r) & = & -\beta \phi_1(r) \hfill \mbox{ for $r > \sigma$ } 
\label{ORPAout} \\
\Delta h(r) & = & ~~0  \hfill           \mbox{ ~~~~~~~~for $r < \sigma$ }. 
\label{optimization}
\end{eqnarray}

These equations  impose, up to the finite crossover distance $\sigma$, 
the matching  of the asymptotic long range  
behavior of $\Delta c(r) $ (eq. (\ref{ORPAout})) and the condition that the approximation 
would not 
modify the pair correlation function at short distances (eq. (\ref{optimization})). 
Due to the 
presence of relation (\ref{SRPA}) one could use as 
independent variable either the values of $\Delta h(r)$ at distances 
beyond $\sigma$ or, more conveniently, the values of the function 
$\chi (r) = \Delta c(r)$
for $r < \sigma$ ($\chi (r) = 0$ for $r>0$) .

It is easy to show that equation (\ref{optimization})
corresponds to the extremum
condition for the following functional \cite{WCA,Olivares}
of $ \chi(r)$:

\begin{equation}
F_{RING} [ \chi (r) ] = \frac{1}{2 (2 \pi)^3  \rho } \int
{ \rm d }^3 { \bf q } 
\{ log [ 1 +  S_0(q) p(q) ] - p (q) S_0(q) \}  \label{FORPA}
\end{equation}
where $p(q) = \rho ( \beta  \hat \phi_{1}(q) -  \hat{\chi} (q) ) $.
In a diagrammatic treatment, $F_{RING}$ would correspond to the sum 
of {\it ring}-like diagrams and, as it is well known, for $\chi(r) = 0$
corresponds to the 
random phase approximation which usually violates the core condition
(\ref{optimization}). The ORPA enforces such a condition.

Indeed, by taking the functional derivative with respect to 
$ \Delta \hat{c}(q)$
we have:

\begin{equation}
{ \frac{ \delta F_{RING} }{ \delta  \Delta \hat{c}(q) } } = { \frac{ \rho }
{(2\pi)^3} \Delta \hat{h}(q)  } \label{gradORPA}
\end{equation}
and Fourier-transforming to the r-space we get for all the values of $r$:

\begin{equation}
 { \frac{ \delta F_{RING} }{ \delta \Delta c(r) } } = \Delta h(r).
\label{deltah}
\end{equation}

For $0<r<\sigma$, equation (\ref{deltah}) becomes an integral equation for the
unknown $\chi(r)$, different from zero only in such a region:

\begin{equation}
 { \frac{ \delta F_{RING} }{ \delta \chi(r) } } = 0 \hfill           \mbox{
 ~~~~~~~~for $r < \sigma$ }.
\label{optimum}
\end{equation}

Thus, equation (\ref{optimum}) is equivalent to imposing an extremum
condition on $F_{RING}$
with respect to variations of $\Delta c(r) = \chi(r) $ (inside $\sigma$). 
If the reference
potential is such that $g_0(r)$ inside $\sigma$ is zero, we see that the
extremum condition is equivalent to the physical requirement that the size of
the exclusion hole of the reference system is preserved by the
perturbation.

It is quite easy to verify that the solution of the equation 
(\ref{optimization}), 
provided it exists,
is actually unique and corresponds to the maximum of the ORPA
generating functional\cite{ORPA1} (\ref{FORPA}).

To complete the description of the system, an explicit prescription for the
reference system pair correlation function $g_0(r)$ is required. In the usual
approach to ORPA, the choice of 
$g_0(r)$ is treated as a separate step. Here we prefer to define a ``total''
generating functional from which the full ORPA $g(r)$ is derived.
Actually we can introduce two functionals having both $F_{RING}$ as generator 
of the ``ORPA'' contribution to the pair correlations  
and differing in the resulting $g_0$.

For a homogeneous liquid interacting through a pair potential $\phi$,
the Helmoltz free energy per particle $F$ can be considered\cite{HMD}  a 
functional of $\phi(r)$ as well
as a functional of the function $e(r)= e^{-\beta \phi(r)}$.
It is easy to show that
\begin{equation}
\frac{ \delta F}{ \delta \phi(r) } = \frac{2}{\rho} g(r)
\label{gfromf}
\end{equation} and that
\begin{equation}
\frac{ \delta F}{ \delta e(r) } = \frac{2}{\rho} y(r),
\label{yfromf}
\end{equation}
where $y(r)$ is the so-called cavity correlation function 
$y(r)=g(r)e^{\phi(r)}$.
We introduce two functionals --- ${\cal{F}}_{GB}[ \Delta \phi ]$ and 
${\cal{F}}_{WCA} [ e ]$ ---
as follows.

\begin{equation}
{\cal{F}}_{GB} [ \Delta \phi ] = \frac{\rho}{2} \int {\rm d} {\bf r}
g_{0}(r) \beta \Delta \phi(r) - \frac{1}{2 \rho} \biggr (\frac{1}{2 \pi} \biggl)^{3}
\int {\rm d} {\bf  q} \biggl [ p(q)S_{0}(q) -log (1 + p(q)S_{0}(q) )\biggr], 
\label{OGB}
\end{equation}

\begin{equation}
{\cal{F}}_{WCA} [ e ] = \frac{\rho}{2} \int {\rm d} {\bf r}
y_{0}(r)  \Delta e(r) - \frac{1}{2 \rho} \biggr (\frac{1}{2 \pi} \biggl)^{3}
\int {\rm d} {\bf  q} \biggl [ p(q)S_{0}(q) -log (1 + p(q)S_{0}(q) \biggr].     \label{OWCA}
\end{equation}

By functional differentiation of ${\cal{F}}_{GB}$ and ${\cal{F}}_{WCA}$
with respect to $\phi(r)$ and $e(r)$ respectively, we get

\begin{equation}
g = g_0 + \Delta h(r)
\label{gGB}
\end{equation}
and
\begin{equation}
y = y_0 + e^{\beta \phi_1} \Delta h(r).
\end{equation}
From the last equation we get immediately the $g(r)$ resulting from 
${\cal{F}}_{WCA}$
as
\begin{equation}
g = g_0e^{-\Delta \phi(r)} + \Delta h(r).
\label{gWCA}
\end{equation}

Thus, the  functionals $ {\cal{F}}_{GB}$ and ${\cal{F}}_{WCA}$ 
are  such that the deviation 
from the 
reference system pair correlation function is always given by the $ORPA$
approximation $\Delta h(r)$ (eqn. \ref{SRPA}), but
the reference system pair correlation function 
is $g_0$ in  one case and $ y_0 e^{-\beta \phi_1} $ in the other case. 
Due to the
form of the  reference system pair correlation functions and the corresponding
generating functionals, we refer to the
former approximation as the Gibbs-Bogoliubov ORPA (GB-ORPA) and to the 
latter as the
Weeks-Chandler-Andersen ORPA (WCA-ORPA). 

Notice that at this level the two functionals have been introduced just as
generating functionals for the pair correlation functions and we are not allowed yet 
to identify the values of the two
functionals at the extremum with the Helmoltz free energy.

\section{ORPA for continuous potentials}

For a general value of the parameter $\sigma$, the solution $\chi(r)$ of equation
(\ref{optimum})  and the resulting $\Delta h(r)$ are
discontinuous at $\sigma$ no matter if the reference system potential is
continuous or not. While such a discontinuity looks relatively 
harmless if the  reference $g_0(r)$ has a hard core of diameter $\sigma$,
a discontinuity in $\Delta h(r)$ at  would be spurious in connection
with a continuous reference system. 

For a similar problem, occurring  in the case of 
the mean spherical approximation (MSA), a
satisfactory solution was found\cite{gillan} by determining $\sigma$ in such a 
way that the
resulting correlation functions were continuous at $\sigma$.
Here, we  can similarly impose the continuity of $\Delta c(r)$ (or equivalently
$\Delta h(r)$) at $r=\sigma$.
Thus , we add the condition
\begin{equation}
\Delta h( \sigma^+ ) = 0
\label{continuity}
\end{equation}
as additional equation for $\sigma$.

Moreover, still in analogy with the MSA case, we can prove that the continuity
condition at  $\sigma$ is equivalent to an extremum condition of the ORPA
functional (\ref{FORPA}) as a function of $\sigma$.

As shown in appendix A we have
\begin{equation}
\partial F_{ORPA} / \partial \sigma = 2\pi \sigma^2\Delta\chi^2(\sigma^-).
\label{deriv}
\end{equation}
Thus, the continuity condition on the correlation functions 
implies that the GB-ORPA and WCA-ORPA functionals  have an extremum (inflection point) at $\sigma$.
As we will discuss in the next section, 
this extremum condition is
also the clue for a thermodynamic consistent theory.

Here we just notice that there is a manyfold of  solutions of equation 
(\ref{continuity}).
However a lower limit for
$\sigma$ is given by the size of the excluded volume region of the reference
system. That is, the region such that

\begin{equation}
g(r)  \approx  0
\end{equation}

A choice of $\sigma$ smaller than the reference system exclusion hole would
result again in an unphysical ORPA $g(r)$. 
On the other hand, since ${\cal{F}}_{WCA}$ and ${\cal{F}}_{GB}$ are 
increasing functions of $\sigma$ (eqn. \ref{deriv}), the minimum value of
will be achieved for the first value of $\sigma$ larger than the
reference system exclusion hole. Moreover, increasing $\sigma$, the size of the
ORPA correction to the reference system thermodynamics and correlations rapidly
decreases.

\section{Generating functionals and thermodynamic consistence}

Now we are in the position to discuss the thermodynamic interpretation of the
functionals ${\cal{F}}_{GB}$ and ${\cal{F}}_{WCA}$ and the specific issue of the 
thermodynamic consistency.

It is well known that approximate integral equation theories for the correlation
functions show quantitative violations of fundamental thermodynamic equalities.
In particular, here we are concerned with the equalities generated by the
identification of the generating functional  with the Helmoltz free energy
per particle  $f$. The most obvious of such equalities  is the equality between 
the
pressure $p$ obtained from the free energy per particle $f$,

\begin{equation}
\frac{ \beta p}{ \rho } = \rho \frac{ \partial{(\beta f)}}{ \rho }
\end{equation}
and that found through the virial theorem,
\begin{equation}
\frac{ \beta p}{ \rho } =  1 - \frac{1}{6} \rho \int g(r) r \beta \phi'(r) 
{\rm d}
{\bf r}
\end{equation}

A necessary condition to ensure that a functional $F [ \phi ]$ is actually a
free energy functional is the validity of equation (\ref{gfromf})
(or (\ref{yfromf}))\cite{free1}.

Such a condition would be fulfilled by the functionals defined in equations
(\ref{OGB}) and (\ref{OWCA}) if  the
dependence of such functionals on all the parameters of the reference
system, on $\sigma$ (say $a_{i}$) and on
$ \Delta \chi$ vanishes. Then we have to satisfy the following equations:
\begin{equation}
{\partial F \over \partial a_i}   = 0 \label{ref}
\end{equation}

\begin{equation}
{\partial F      \over \partial \sigma}   = 0
\label{contin}
\end{equation}

\begin{equation}
{\delta F \over \delta  \chi(r) }   = 0
\label{ORPAin}
\end{equation}

Eqn (\ref{ORPAin}) corresponds to the ORPA formula [7] while eqn (\ref{ref}) is a way of
determining the reference system parameters. 
Eqn. (\ref{contin}), as we have  shown in the previous section
is also related  to  the continuity of the resulting correlation functions.
Therefore, in order to have thermodynamic consistency we have to ensure that the
functional would be extremum with respect to variations of {\em all} 
the parameters.
An analogous requirement for the choice of the reference system in connection 
with the modified hypernetted chain approximation (MHNC)  was derived by 
Lado et al.\cite{ladovar}. Even closer to the present problem is the analysis of
the choice  of the reference system within the WCA perturbation theory provided
by Lado\cite{ladowca}.

Different choices of the reference system parameters are certainly conceivable and
actually this is the existing situation. It is not easy to anticipate what is 
the best
choice for all possible systems and a final assessment should be left to explicit
numerical investigations. However, here we can notice  that only the choices
corresponding to extrema of the generating functionals or choices completely
independent of the thermodynamic state would ensure the free energy nature of the
generating functionals and then, as a consequence, the consistence of the energy and
virial routes to the equation of state.

\section{Conclusions}

In the present paper we have rephrased the ORPA in the language of the 
generating functionals for the pair correlation function.
In this way we could easily obtain three main results:

\begin{itemize}

\item[1)] we can derive  from a unified treatment the two prescription 
for the reference system $g(r)$ ( equations (\ref{gGB}) and (\ref{gWCA}) ) 
present in the literature;

\item[2)] we can show how the ORPA  can be extended to deal 
with continuous reference system interactions, potentially increasing 
the range of applicability of this approximation;

\item[3)] we show that the closure equations, the removal of the 
discontinuity in the resulting pair correlations and  the 
identification of the generating functionals with the Helmoltz free 
energy can be reduced to the unique and unifying requirement of a variational 
principle on the functionals with respect to all the independent 
variables and parameters.

\end{itemize}

The theory presented in this paper provides a general scheme 
corresponding to many possible choices for the individual ingredients 
of the ORPA. Actually, depending on the reference system and on the flavor of
the ORPA (GB
or WCA), we have 
introduced  different possibilities. For this reason we postpone 
detailed  numerical investigations to the 
application of the approximation to specific problems.

Taking into account the already satisfactory level of accuracy of the 
standard implementations of the ORPA, and judging from  preliminar 
calculations, we can anticipate a good quality of the resulting 
numerical results. In particular thermodynamical investigations could 
now benefit from the clarified status of thermodynamic consistency in 
the ORPA. In this respect, we believe that the ORPA could  play an 
important role as one of the best candidate for the investigation of 
the fluid phase  diagrams.

\vskip 2truecm
{\bf Acknowledgements}
O.A, F.M and Z.B wish to thank the International Atomic Energy Agency and UNESCO
for grant allowing their articipation in the Condensed Matter activities at the
International Centre for Theoretical Physics in Trieste.

\smallskip

\newpage
\centerline{\bf APPENDIX A}
The proof of eqn. (\ref{deriv}) is given as follows: \\
Let
$$p(q) = \beta \rho  \hat \phi_1 (q) - \rho \hat \chi(q). \eqno(A1)$$

Since the only dependence of ${\cal{F}}_{WCA}$ or  ${\cal{F}}_{WCA}$ on 
$\sigma$ is through $\chi$, the derivative of 
$F_{RING}$  (eqn. (\ref{FORPA})), we have to evaluate

$${\partial F_{RING} \over \partial \sigma}= \frac{1}{2\rho}\biggl (\frac{1}{2 \pi}\biggr )^3
\int {\rm d} {\bf q}
\biggl [S_o(q)-S(q) \biggr ]{\partial p(q) \over \partial\sigma} 
= -\frac{1}{2}\biggl (\frac{1}{2 \pi}\biggr )^{3} \int \rm d{\bf q} 
\Delta \hat h(q){\partial \hat \chi (q)
\over \partial\sigma}. \eqno(A3)$$
Now, taking into account the finite support of $\chi(r)$,
$${\partial  \hat \chi (q) \over \partial \sigma}= {4\pi \over q} \sigma 
\chi (\sigma^-)sin (q\sigma) + \int_{0}^{\sigma} r { \partial  \chi (\sigma) 
\over
\partial \sigma} {4\pi \over q} sin (qr) dr. $$
By using Parseval's equality, eqn. (A3) becomes:
$${\partial F_{RING} \over \partial \sigma}=-\frac{1}{2}\int {\rm d} {\bf  r} \Delta h(r) {\partial 
\chi (r) \over \partial \sigma} - \frac{1}{2} \biggl ( \frac{1}{2 \pi} 
\biggr)^{3}
(4\pi)^2 \int_{0}^{\infty}  q \Delta \hat h(q) \sigma  \chi
(\sigma) sin (q\sigma) dq \eqno(A4)$$
the first term in eqn. (A4) is zero because when $\Delta h \neq 0$, the other
term is zero and the reverse also follows. Eqn (A4) eventually reduces to
$${\partial F_{RING} \over \partial \sigma} = -2\pi\sigma^2\chi (\sigma^-) \Delta h(\sigma^+)= -2\pi\sigma^2
\chi^2(\sigma^-) \eqno(A5)$$
giving eqn. (\ref{deriv})) when we take into account that $F_{RING}$ appears in
eqns. (\ref{OGB}) and (\ref{OWCA}) with a negative sign.

\end{document}